\documentclass[12pt]{article}

\def\be{\begin{equation}}
\def\ee{\end{equation}}
\def\ba{\begin{eqnarray}}
\def\ea{\end{eqnarray}}
\def\half{{1 \over 2}}
\def\x{{\bf x}}
\def\tt{\tilde{t}}
\def\tz{\tilde{z}}
\input epsf
\begin{document}
\title{Relaxation of the cosmological constant in a movable brane world}
\author{S. Khlebnikov \vspace{0.1in} \\
{\it \normalsize Department of Physics, Purdue University, West Lafayette, 
IN 47907, USA}}
\date{July 2002}
\maketitle
\begin{abstract}
We present numerical evidence that a domain wall in a background with 
varying vacuum energy density acquires velocity in 
the direction of decreasing Hubble parameter. This should lead to at least
a partial relaxation of the cosmological constant on the wall.\\
\hspace*{3.9in} hep-th/0207258
\end{abstract}
Brane-world scenarios, according to which we live on a domain wall in 
a higher-dimensional
space \cite{Rub&Shap}, 
open new avenues for thinking about the cosmological constant.
Here, we want to explore feasibility of one particular mechanism that
may help the ``constant'' to relax to a small value.

The effective cosmological constant seen by an observer on the wall is 
determined by the local value of the Hubble parameter, $H$.
It has long been known \cite{infl_wall} that a domain wall in a space
with {\em zero} vacuum energy will inflate, i.e. $H$
will be nonzero. To achieve $H=0$, one may introduce a {\em negative}
vacuum energy in the bulk, with a value tuned precisely to compensate for
the effect of the wall tension \cite{Randall&Sundrum}.
Alternatively, one can introduce a scalar field with a dilaton-like
coupling to the brane, to obtain static solutions for more generic values 
of the tension \cite{self-tune}. Or, with more 
than two external dimensions, one may keep the bulk vacuum energy 
zero but assume that gravity in the bulk is so weak that the Hubble parameter
induced by the wall tension is small enough to be compatible with observations
\cite{Dvali&al}. 

Here, we want to consider a different idea. If the bulk vacuum energy
varies along an extra dimension, the wall may be moving in such a way
that the Hubble parameter on it starts
at a large positive value but then approaches zero---or even
crosses zero and becomes negative. In this way, the wall will move towards
a region where two large contributions to $H$
(one from the bulk energy, and the other from the wall tension) nearly cancel.
In other words, a finely tuned solution similar to that of 
ref. \cite{Randall&Sundrum} will be reached automatically in the course
of the evolution. If the cancellation persists for some time,
it may be possible to identify that period of time with
the standard cosmology.\footnote{
Similarly, $H$ can change sign if the wall
itself is static but the bulk vacuum energy or the wall tension change 
in time. In this case, however, it may be more difficult to make the time
when $H$ is close to zero long enough to include the entire known
cosmological history.}

How closely $H$ can get to zero in this scenario will depend on 
the complicated dynamics occurring when the above two contributions become
of the same order. In this paper, we address an earlier, simpler stage of
the evolution, when the effect of the wall tension is still much smaller
than the effect of the bulk vacuum energy. We have tried to obtain 
numerically geometries in which $H$ changes little with time but varies
significantly in space, and see which way a test domain wall will move 
in such a background. Our main result is that, in the backgrounds we were 
able to obtain, the wall quite generically acquires velocity in the 
direction of {\em decreasing} Hubble parameter, i.e. towards regions 
that are ``less inflationary''. Although the actual displacements that we
have been able to detect in the lifetime of our simulation are quite small,
it is encouraging that the final velocity is a sizable fraction of 
the speed of light.

A suitable smooth background can be formed, for example, by another domain 
wall, which is much thicker than ours. The
difference in the thickness may be due to a difference in masses, or simply
in initial profiles, of the scalar fields forming the walls.
Towards the end of the paper, we will speculate that there is also a more 
``economical'' possibility, wherein a configuration of one domain wall in 
the background of another is realized with a single field 
in the context of topological inflation \cite{top_infl}. 
For most of the paper, though, 
we will prefer to think about these two walls as being due to two different 
fields.

In the remainder of the paper, we (i) describe some a priori guesses 
about geometries induced by a space-dependent vacuum energy and 
(ii) present a sample of the numerical results.

We prefer to use isotropic coordinates, in which the metric has the form
\be
ds^2 = -N^2(z,t) dt^2 + \psi^4(z,t) [ dz^2 + d\x^2 ] \; ;
\label{ds2}
\ee
$N$ is called lapse, and $\psi$---conformal factor. This choice of
coordinates was adopted from our numerical studies of post-inflationary 
universe \cite{FK}, and the numerical algorithm we use here follows closely 
the one used in that work. We allow for the total of $D > 1$ spatial 
coordinates, but the metric depends only on one of them---the fifth 
coordinate $z$. Thus, $d\x^2$ is the flat Euclidean metric in 
the remaining $(D-1)$ dimensional space.

For a preliminary discussion, it will suffice to use the energy
constraint
\be
\frac{\kappa^2 C_E}{D-1} \equiv \frac{\kappa^2 E}{D-1} - \half D H^2 
+ 2 \frac{\psi''}{\psi^5} + 2(D-3) \frac{(\psi')^2}{\psi^6} 
= 0 \; ,
\label{ec}
\ee
and the equation for the lapse
\be
\frac{N''}{N} - \frac{4\psi' N'}{\psi N} + 2(D-2) \frac{\psi''}{\psi}
- 6(D-2) \left( \frac{\psi'}{\psi} \right)^2 = 
- \kappa^2 \sum_n (\phi'_n)^2 \; .
\label{eqN}
\ee
Here $E$ is the energy density (defined below), and
\be
H = \frac{2\dot{\psi}}{N\psi} 
\label{eqpsi}
\ee
is the local Hubble parameter. The sum on the right-hand side of (\ref{eqN})
is over all scalar fields $\phi_n$ of the model; a prime denotes derivative
with respect to $z$, and a dot---with respect to $t$; $\kappa^2 = 8\pi G$;
$G$ is bulk Newton's constant.

We first look for purely vacuum solutions with a constant energy density.
In this case, the right-hand side of (\ref{eqN}) is zero, while $E$ 
in (\ref{ec}) equals the cosmological constant: $E=\Lambda$. The equations
have the following solution:
\ba
\psi & = & \frac{A}{\sqrt{z -\alpha t}}  \; , \label{solpsi} \\
N & = & \psi^2  \; ,  \label{solN}
\ea
where $A$ and $\alpha$ are constants. The Hubble parameter (\ref{eqpsi}) is
constant: $H = \alpha / A^2$, so whenever $\alpha > 0$ the solution is
inflationary. Note that for $\alpha >0$, the solution is well-defined in
the half-space $z > 0$ for any $t < 0$. This is appropriate, since according
to (\ref{solN}), $t$ is the conformal time, which for inflationary solutions
is limited to negative values.

Eq. (\ref{eqN}) is satisfied by (\ref{solpsi})--(\ref{solN}) identically,
but the energy constraint (\ref{ec}) imposes a relation between
the parameters:
\be
\alpha^2 - 1 = \frac{2\kappa^2 \Lambda}{D(D-1)} A^4 \; .
\label{rel}
\ee
This is the same relation as obtained for a class of solutions in ref.
\cite{Kaloper}, so ours should be the same solutions except written in
different coordinates. In particular,
for $\Lambda = 0$, we have $\alpha = \pm 1$ and any $A$. ($A$ will be
determined by the junction condition on the wall: e.g. for a static thin 
wall with the geometry symmetrical about it,
$A^{-2} = \kappa^2 \sigma / 2(D-1)$, where $\sigma$ is the wall tension.)
The $\alpha=1$ solution is the inflationary solution
of ref. \cite{infl_wall} in different coordinates.\footnote{The explicit
relation between our coordinates and those used in ref. \cite{infl_wall}
and called here $\tz$ and $\tt$ is:
$z=\exp(-\tt)\sinh \tz$, $t=-\exp(-\tt)\cosh \tz$, 
all in units of the constant Hubble distance $H^{-1}$.} 
To obtain $\alpha = 0$ (and consequently $H = 0$),
we need to have $\Lambda = -D(D-1)/ 2\kappa^2 A^4 $. This is the anti-de Sitter
solution of ref. \cite{Randall&Sundrum}.

Now, when $\Lambda$ is not exactly a constant but is slowly varying, 
we expect the geometry to interpolate between faster expanding regions 
with larger $\Lambda$ and slower expanding, or perhaps even contracting, 
regions with smaller $\Lambda$. Geometries interpolating between de Sitter 
and and anti-de Sitter spaces can presumably be obtained by analytical 
continuation of certain Euclidean instantons. Such instantons, 
containing a de Sitter region sandwiched between two anti-de Sitter regions, 
have been considered in
ref. \cite{HMS} in connection with the problem of counting de Sitter 
microstates. Note that for every positive $\alpha$ (expanding) solution,
eq. (\ref{rel}) has a negative $\alpha$ (contracting) solution. Such contacting
regions are ubiquitous in our numerical simulations. We do not know, however,
if they can obtained by analytical continuation from any instantons.

To study real-time evolution that may possibly lead to such interpolating 
geometries, we will need the rest of Einstein's
equation. As in ref. \cite{FK}, we write these in the Hamiltonian form.
The Hamiltonian pair to eq. (\ref{eqpsi}) is
\be
\frac{\dot{H}}{N} = - \frac{D}{2} H^2 + 2(D-2) \frac{(\psi')^2}{\psi^6}
+ \frac{2\psi' N'}{\psi^5 N} 
+ \frac{\kappa^2}{D-1} \left( 2V - E + w C_E \right) \; .
\label{eqH}
\ee
(Although we did not emphasize that above, for a constant $E=V=\Lambda$
eq. (\ref{eqH}) is also satisfied by the solution 
(\ref{solpsi})--(\ref{solN}).)
Equations for the fields are
\ba
\dot{\phi}_n & = & N \frac{\pi_n}{\psi^{2D}} \; , \label{eqphi} \\
\dot{\pi}_n & = & \left( \psi^{2D-4} N \phi'_n \right)' 
- N \psi^{2D} \frac{\partial V}{\partial \phi_n} \; , \label{eqpi}
\ea
The energy density is
\be
E = \sum_n 
\left( \frac{\pi_n^2}{2\psi^{4D}} + \frac{(\phi_n')^2}{2\psi^4} \right)
+ V(\phi) \; ;
\label{E}
\ee
$V$ is the potential of the fields. The energy constraint $C_E$
can be added to the right-hand side of (\ref{eqH}) with any coefficient 
$w$ (however, for numerical studies some choices may be better than
others \cite{FK}). In what follows, we use $w=1$.

The Hamiltonian structure of the equations allows us to use a leap-frog
algorithm, in which the coordinates $\phi_n$ and $\psi$ are defined
at full time steps, while the momenta $\pi_n$ and $H$ are defined at 
half-steps. The lapse does not have a Hamiltonian pair; as in \cite{FK},
we update it at full steps using eq. (\ref{eqN}) and extrapolate it to
half-steps. The energy constraint (\ref{ec}) is not used for the evolution
but is monitored to provide an idea of how accurate the results are.
In addition, we have monitored the momentum constraint
\be
\kappa^2 \psi^2 C_M \equiv \frac{\kappa^2}{\psi^{2D}} \sum_n \pi_n \phi'_n +
(D-1) H' = 0
\label{mc}
\ee
and have found that for the time intervals shown in the plots both $C_E$ 
and $C_M$ stay close to zero to a satisfactory degree.

To start numerical evolution, we need to solve the constraints (\ref{ec})
and (\ref{mc}) at the initial time slice. The momentum constraint (\ref{mc}) 
is solved trivially with $H = {\rm const}$ and the initial choice of the fields 
such that for each $n$ either $\pi_n$ or $\phi'_n$ vanish. There is some
advantage in being able to choose a suitable initial profile of $\psi$
by hand. To make our choice compatible with the energy constraint (\ref{ec}), 
we introduce a massive ``radiation'' field which is initially zero and
whose initial momentum is determined by solving eq. (\ref{ec}).
(This is similar to the method used in ref. \cite{GP} to study the onset of
inflation in inhomogeneous spacetimes.) We call this field ``radiation'' only
to distinguish it from the fields forming the domain walls; its mass is in 
fact the largest mass in the problem. The ``radiation'' energy
redshifts away in the course of the evolution (of course, more so in
regions that are inflationary).

We thus have three fields in our simulation: $\phi_0$ that forms the thick 
background wall,
$\phi_1$ that forms the movable thin wall, and $\phi_2$, the massive
``radiation'' field.  We keep the thick wall centered at $z=0$ by choosing 
the boundary condition $\phi_0(0,t) = 0$. For the potential, we use
\be
V = \sum_{n=0,1} \left( -\half \mu_n^2 \phi_n^2 + {1\over 4} \lambda_n 
\phi_n^4 \right) + \frac{\mu_1^4}{4\lambda_1} + \half M^2 \phi_2^2 + 
\Lambda \; ,
\label{V}
\ee
where $\Lambda$ is a constant---it is the vacuum energy at $z=0$.
A nontrivial $z$-dependence of vacuum energy is due to variation of the
field $\phi_0$. With potential (\ref{V}),
the only interaction between different fields is gravitational.

A word about the choice of parameters. We measure times and lengths in units
of $\mu_0^{-1}$ and energy densities in units of $\mu_0^4/ \lambda_0$, 
so that we can
set $\mu_0 =1$ and $\lambda_0 = 1$. The results below are for $\mu_1 = 2$,
$\lambda_1 = 2000$, $M=250$, $\kappa^2 = 37.5$ 
(this dimensionless $\kappa^2$ is $\kappa^2 \mu_0^2 / \lambda_0$
in the original units), $\Lambda = 0.16$, and $D=4$ (i.e. one extra dimension).

We want the background to be sufficiently smooth, so that the interior of
the thick wall---a region near $z=0$---can inflate. This type of inflation
is sometimes called topological inflation \cite{top_infl}. 
So, we choose the initial profile of $\phi_0$ in the form
\be
\phi_0(z, t=0) = \frac{z}{\sqrt{z^2 + \ell^2}} \; ,
\label{phi0}
\ee
with the spatial scale $\ell$ much larger than the wall's 
``natural'' thickness: $\psi^2 \ell \gg \mu_0^{-1}$. 
The plots below are for
$\ell = L = 50$, where $L$ is the total length of the integration
region: $z \in [0,L]$, from a grid with the total of 2049 points.

Another necessary condition for topological inflation is that near $z=0$
the Hubble parameter supported by the vacuum energy of $\phi_0$ is large
enough, in comparison with $\mu_0$ \cite{top_infl}.
With the above values of the parameters, this
Hubble parameter is $H_0 = (\kappa^2 \Lambda / 6)^{1/2} = 1$. There have been
numerical determinations of the critical ratios $H_0/\mu_0$ for
which the interiors of various types of topological defects will 
inflate in three spatial dimensions \cite{num_stu}. We have not attempted
a similarly detailed study for $D=4$, but have observed large amounts
of inflation for the above value of $H_0/\mu_0 = 1$.

In comparison, gravitational effects due to the wall formed by $\phi_1$
are, by our choice of the parameters, much smaller. 
A convenient measure of these effects is 
the expansion rate the wall {\em would} have if it were inflating
in empty space with zero bulk cosmological constant: 
$H_1 = \kappa^2 \sigma_1/ 2(D-1)$, where 
$\sigma_1 \sim \mu_1^3/ \lambda_1$ is the wall tension. For the above
values of the parameters, this gives 
$\sigma_1\sim 4\times 10^{-3}$. This estimate is well born out 
numerically,\footnote{Variation of the tension with time is relatively
small, which implies that the wall formed by $\phi_1$ maintains more or
less constant physical (as opposed to comoving) thickness.} so
we use $H_1=0.03$. Because $H_1 \ll H_0$,
the wall formed by $\phi_1$ has only a minor effect on the geometry
in the rapidly expanding region near $z = 0$. 
On the other hand, when the wall is 
in a region where the expansion rate is of order $H_1$, or smaller, we
expect strong gravitational effects due to the wall's tension.
As we have already mentioned, such effects are
likely to determine if the present scenario leads to
a complete solution to the cosmological 
constant problem, but they lie outside the scope of this initial study.

\begin{figure}
\leavevmode\epsfysize=3.0in \epsfbox{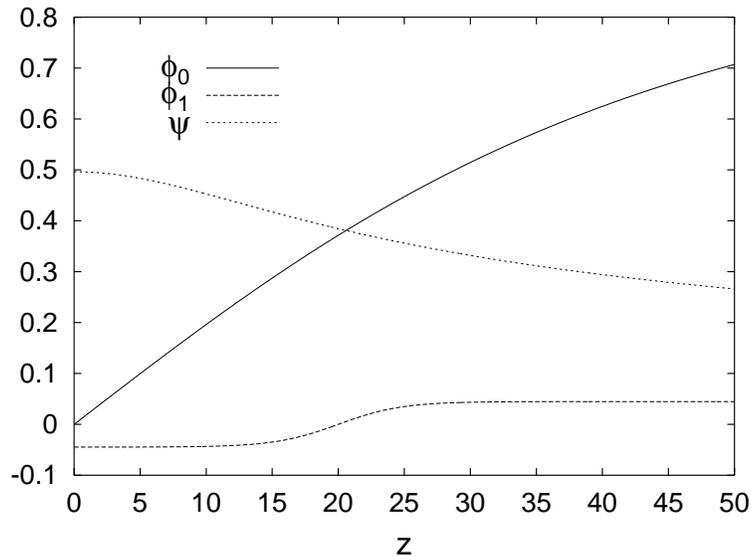}
\vspace*{0.2in}
\caption{Initial profiles of the two domain walls and of the conformal
factor $\psi$.
}
\label{fig:init}
\end{figure}

In Fig. \ref{fig:init}, 
we show the initial profiles of $\phi_0$, $\phi_1$, and $\psi$
for a typical initial position of the second wall, $z_1 = 20$.

\begin{figure}
\leavevmode\epsfysize=3.0in \epsfbox{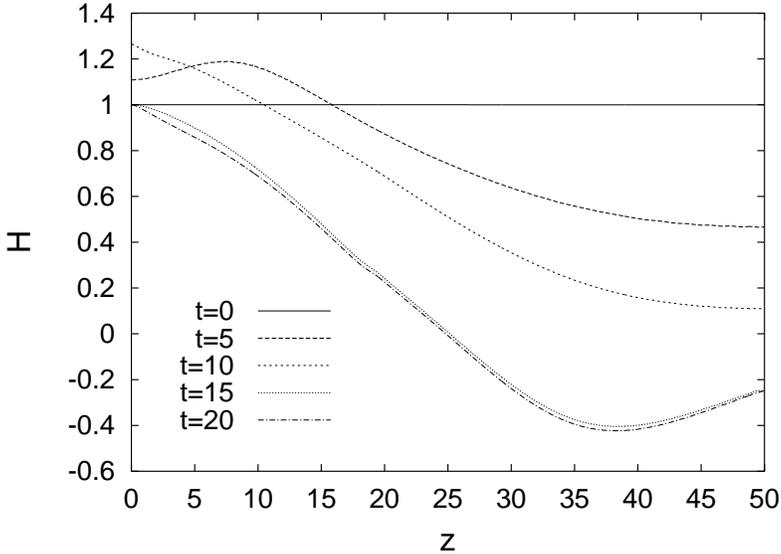}
\vspace*{0.2in}
\caption{Profiles of the local Hubble parameter at several moments 
of time.}
\label{fig:H}
\end{figure}

Now, let us see if an interpolating geometry of interest to us is
reached in the course of numerical evolution. Evidence for that is shown
in Fig. \ref{fig:H}, where we plot profiles of the Hubble parameter $H$
at several moments of time. We see that, after a transient behavior,
$H$ acquires a nearly static profile, in which it has positive values
near $z=0$ (where the vacuum energy is the largest) and negative values
at large $z$.

\begin{figure}
\leavevmode\epsfysize=3.0in \epsfbox{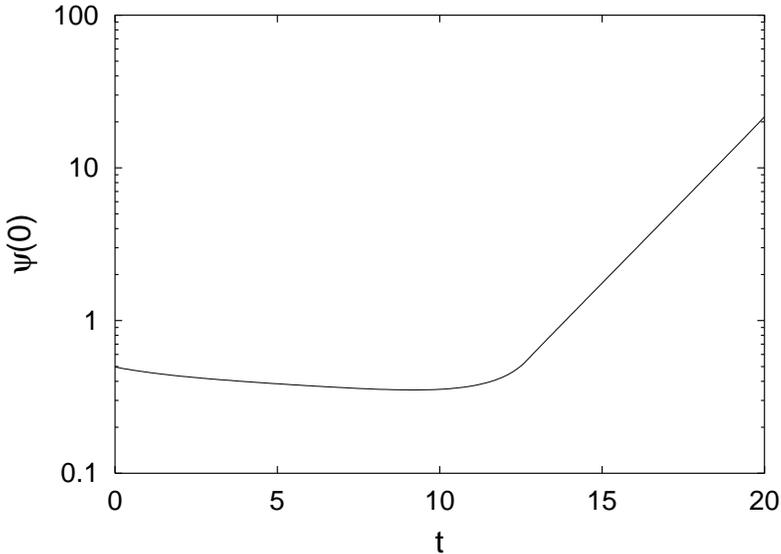}
\vspace*{0.2in}
\caption{Inflation at the origin: conformal factor at $z=0$ as a function 
of time.}
\label{fig:psi}
\end{figure}

To estimate the amount of inflation we obtain near $z=0$, we plot in 
Fig. \ref{fig:psi} the value of the conformal factor $\psi$ at $z=0$ 
as a function of time. We see an exponential growth by a factor over 40
between $t=12.5$ and $t=20$. This corresponds to the size of the universe 
growing by a factor over 1600.\footnote{To properly interpret the plot,
one needs to know also the time-dependence of the lapse $N$ at $z=0$, since 
it is $N(z,t) dt$ that determines the increment of cosmic time at 
location $z$. As it happens, after $N$ reaches the value of 1 at $z=0$ in our
simulation, we keep it equal to 1 there at all subsequent times (which 
is always possible due to the remaining gauge freedom). For Fig. \ref{fig:psi}, 
this resetting of the clock rate took place  at $t\approx 12.7$; 
after that, time
$t$ coincides (up to a constant) with the cosmic time at $z=0$.}

We define velocity of the domain wall formed by $\phi_1$ as the ratio
of the momentum of $\phi_1$ to its energy, both taken per unit transverse 
area:
\be
v = -\frac{\int \pi_1 \phi'_1 \psi^{-2D} dz}{\int E_1 \psi^2 dz} \; ,
\label{vel}
\ee
where
\be
E_1 = \frac{\pi_1^2}{2\psi^{4D}} + \frac{(\phi_1')^2}{2\psi^4} 
+ {1\over 4} \lambda_1 \left( \phi_1^2 - \frac{\mu_1^2}{\lambda_1} \right)^2 
\; .
\label{E1}
\ee
(The denominator in (\ref{vel}) divided by the $\gamma$-factor of the wall
is our definition of the wall tension 
$\sigma_1$.) Note that if the profile of $\phi_1$ with
respect to the {\em comoving} coordinate $z$ does not change, 
i.e. $\partial_t \phi_1 =0$,
the definition (\ref{vel}) gives zero. In other words, $v$ as defined by
(\ref{vel}) is insensitive 
to a motion that is due solely to the expansion of the universe: it 
measures velocity of the wall relative to the expanding background.

\begin{figure}
\leavevmode\epsfysize=3.0in \epsfbox{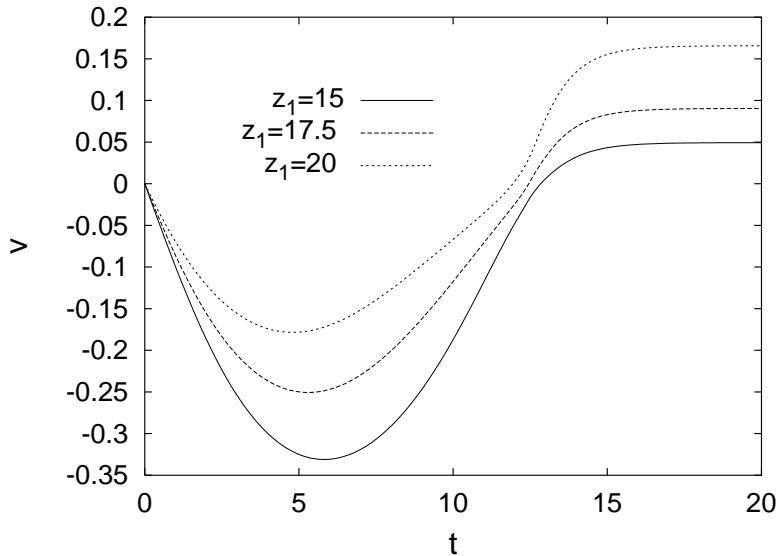}
\vspace*{0.2in}
\caption{Velocity of the domain wall formed by $\phi_1$ as a function of
time, for three different values of the initial position.}
\label{fig:vel}
\end{figure}

In Fig. \ref{fig:vel} we plot the wall's velocity as a function of time,
for the same initial
position of the wall $z_1=20$ that was used for the previous plots, as
well as for two smaller values of $z_1$. We see that $v$
invariably becomes positive at late times, when a geometry with an almost
static profile of the Hubble parameter is reached. 
The final value of the velocity can be a sizable fraction of the speed
of light and is larger for walls that are
further away from the origin.

It may appear from Fig. \ref{fig:vel} that the velocity goes to a 
steady asymptotic value, but we should point out that
the lapse $N(z,t)$ at the locations of the wall drops
precipitously during the late-time evolution. So, the entire stage when
$v$ is close to its final value corresponds only to a small interval 
of cosmic time on the wall. As a consequence, the corresponding
displacements of the wall are tiny. On the other hand, one can argue
that the velocity remains steady over large amounts of expansion 
in the inflating region, cf. Fig. \ref{fig:psi}.

Throughout this paper, we have considered the case when the wall formed
by $\phi_1$ is much ``weaker'', in terms of its gravitational effects, than
the wall formed by $\phi_0$. This does not always have to be so, and, as
we have already mentioned, perhaps the most interesting case with regard
to the cosmological constant problem is when the ``strengths'' of the two 
walls are comparable. In addition, walls of different thicknesses but 
of equal ``strengths'' naturally arise in the context of topological 
inflation \cite{top_infl} with an extra dimension. Let us see how.

The wall formed by $\phi_1$ in our simulations maintains more or less
constant physical thickness, as opposed to typical walls of $\phi_0$
considered in topological inflation. Those latter are produced by 
growth of small fluctuations in the {\em middle} of the inflating 
region, where $\phi_0 \approx 0$, and the vacuum energy 
$E_{\rm vac}$ is much larger than the gradient energy $E_{\rm grad}$
of a new wall (whose thickness at birth is of the order of 
the Hubble distance). So, these walls expand in all directions 
\cite{top_infl}. Analogy with the case 
considered here is much closer for a wall produced at the
{\em fringe} of the inflating region. If a wall finds itself in a region 
with $E_{\rm vac} \sim E_{\rm grad}$, its thickness will not grow,
but its gravitational effects will be
comparable to those of the background. This can be seen by comparing 
the previously introduced parameters $H_0$ and $H_1$. Indeed,
$H_0^2 \sim \kappa^2 E_{\rm vac}$,
while $H_1 \sim \kappa^2 \sigma_1 \sim \kappa^2 E_{\rm grad} H_0^{-1}$, so
$H_1 \sim H_0$. 

Production of new domains walls of $\phi_0$ 
should be observable even in the model considered here, if we were 
able to follow the evolution for a long enough time. In fact, it may well be
that the eventual breakdown of our simulations (at times larger than those 
shown in the plots) is due primarily to the onset of this new, more complex
dynamics. Obtaining good quality data for such a regime will
require further numerical work. We hope to return to this question in 
a future publication.

In conclusion, we have found that numerically solving Einstein's equations 
on one-dimensional lattices allows one to obtain, and experiment with,
geometries in which the vacuum energy density varies along an extra dimension,
and the local value of the Hubble parameter is almost static
but varies from point to point. We have also found that a test domain wall
in such a geometry will move in the direction of decreasing Hubble parameter,
suggesting that this mechanism can lead to at least
a partial relaxation of the cosmological ``constant'' in a brane world. 

This work was supported 
in part by the U.S. Department of Energy through Grant DE-FG02-91ER40681 
(Task B).

\end{document}